\documentclass[twocolumn,showpacs,amsmath,amssymb]{revtex4}

\usepackage{graphicx}

\renewcommand{\vec}[1]{\mbox{\boldmath $#1$}}

\begin{document}

\title{
Iterative solution of a Dirac equation with inverse Hamiltonian method}

\author{K. Hagino}
\affiliation{ 
Department of Physics, Tohoku University, Sendai 980-8578,  Japan} 

\author{Y. Tanimura}
\affiliation{ 
Department of Physics, Tohoku University, Sendai 980-8578,  Japan} 


\begin{abstract}
We solve a singe-particle 
Dirac equation with Woods-Saxon potentials 
using an iterative method in the coordinate space representation. 
By maximizing the expectation value of the 
inverse of the Dirac Hamiltonian, 
this method avoids the variational collapse, in which an iterative solution 
dives into the Dirac sea. 
We demonstrate that this method works efficiently, reproducing 
the exact solutions of the Dirac equation. 
\end{abstract}


\maketitle

The imaginary time method has been successfully 
employed in non-relativistic self-consistent mean-field 
calculations \cite{DFKW80,ev8}. 
The idea of this method is that a function $e^{-H\tau /\hbar}|\psi_0\rangle$ 
converges to the ground state wave function of a Hamiltonian $H$ 
as $\tau\to\infty$ for 
any trial wave function $|\psi_0\rangle$ as long as it is not an 
eigenstate of $H$. 
One can then iteratively seek for the eigenfunctions of $H$ 
in the coordinate space starting from 
an arbitrary wave function $|\psi_0\rangle$. 
This method is suitable particularly 
for self-consistent mean-field calculations in 
a three-dimensional coordinate space, with which an arbitrary shape 
of nuclei can be efficiently described \cite{BFH85}. 

A naiive extension of this method to a relativistic equation, however, 
meets a serious problem. That is, a Hamiltonian $H$ has both the Fermi and 
Dirac seas (see Fig. 1) 
and an iterative solution inevitably 
ends up with a wave function in the 
Dirac sea even if the starting wave function $|\psi_0\rangle$ is in the 
Fermi sea. This problem was recently pointed out in Ref. \cite{ZLM10} 
in the nuclear physics context, although the problem had been well known 
in quantum chemistry under the name of {\it variational collapse} 
\cite{G82,LM82,SH84,HK94,PM95,FFC97,DESV00,K07}. 
In Ref. \cite{ZLM10}, the problem was avoided by applying the imaginary 
time method to the Schr\"odinger-equivalent form of the Dirac equation. 
The same method was used also in Ref. \cite{GM07}. 
Moreover, damped relaxation techniques were employed in Ref. \cite{BSUR89} 
for an iterative solution of a Dirac equation. 

In this paper, we propose a yet another method for iterative solution of a 
Dirac equation, which can be directly applied to the coordinate space 
representation of a relativistic wave function. 
The method is based on the idea of Hill and Krauthauser \cite{HK94} 
to maximize the expectation value of the inverse Hamiltonian. 
This method solves the Dirac equation as it is and thus both the 
upper and lower components of a wave function are automatically obtained. 
Also, in this method, 
states in the Dirac sea can be obtained within the same scheme 
by inverting the direction of ``time'' evolution, while the method 
in Refs. \cite{ZLM10,GM07} requires two different equations for 
the Fermi and Dirac seas. Thus, our method can be 
regarded complementary to 
the method in Refs. \cite{ZLM10,GM07}. 

\begin{figure}[htb]
\includegraphics[clip,scale=0.5]{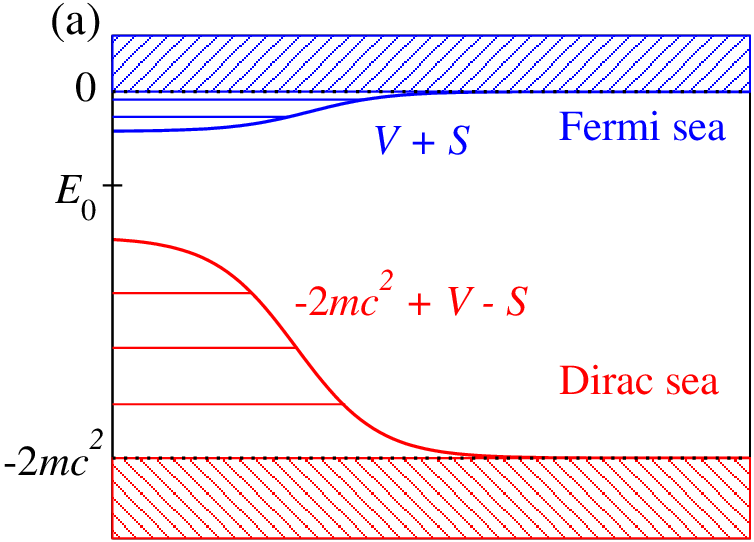}\\
\hspace*{0.35cm}
\includegraphics[clip,scale=0.5]{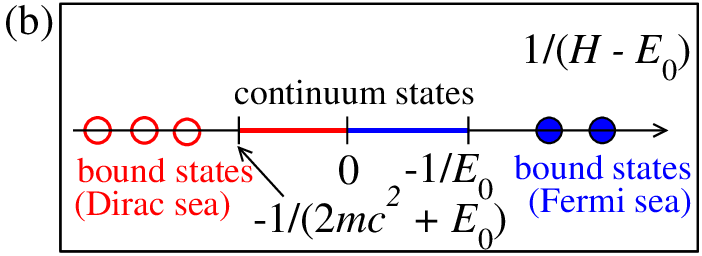}
\caption{(Color online)
(a) A schematic spectrum of a Dirac Hamiltonian $H$. 
$V(r)$ and $S(r)$ are vector and scalar potentials, respectively. 
$E_0$ is an arbitrary constant located between the lowest 
energy state in the Fermi sea and the highest energy state in the Dirac sea.  
The hatched regions indicate the continuum spectra. 
(b) A spectrum of $1/(H-E_0)$ corresponding to Fig. 1(a). 
The filled circles indicate the bound states in the Fermi sea, while the 
open circles indicate the bound states in the Dirac sea. 
The continuum region is denoted by the thick solid line. }
\end{figure}

In our method, the ground state wave function $|\Psi\rangle$ 
of a Dirac Hamiltonian 
$H$ is obtained 
by using the relation
\begin{equation}
|\Psi\rangle \propto \lim_{T\to\infty}e^{\frac{T}{H-E_0}}|\psi_0\rangle,
\label{iteration}
\end{equation}
where $E_0$ is a constant between the lowest eigenvalue in the Fermi sea 
and the highest eigenvalue in the Dirac sea. 
In order to illustrate how this method works, Fig. 1 shows a spectrum of 
$H$ (Fig. 1 (a)) and that of $1/(H-E_0)$ (Fig. 1 (b)) \cite{HK94}. 
In the spectrum of $1/(H-E_0)$ shown in Fig. 1(b), 
the states in the Fermi sea appear 
in the positive region while those in the Dirac sea in the negative 
region. Therefore, as $T\to\infty$ in Eq. (\ref{iteration}), all the 
states in the Dirac sea are damped out. 
Moreover, the lowest energy state in the Fermi sea corresponds to 
the highest point in Fig. 1(b), and only this state survives in 
Eq. (\ref{iteration}) as $T\to\infty$. 

In practice, Eq. (\ref{iteration}) is solved iteratively. That is, from 
the wave function at $T\equiv n\Delta T$, $|\Psi^n\rangle$, the wave function is evolved 
with a small interval $\Delta T$ as 
\begin{equation}
|\Psi^{n+1}\rangle \sim \left(1+\frac{\Delta T}{H-E_0}\right)\, 
|\Psi^n\rangle.
\label{iteration2}
\end{equation}
The inverse of the Hamiltonian is in a familiar form in non-relativistic 
time-dependent Hartree-Fock (TDHF) calculations \cite{KDM77,KM90} 
(see also Ref. \cite{BGS83} for a similar technique in 
relativistic TDHF calculations), and one can evaluate the inverse 
similarly 
to Refs. \cite{KDM77,KM90,BGS83} using the Gauss elimination method. 
We demonstrate it here for spherical potentials. The relativistic wave function 
then takes the form\cite{S67,G97}, 
\begin{equation}
\Psi(\vec{r})=
\frac{1}{r}
\left(
\begin{array}{c}
G(r){\cal Y}_{jlm}(\hat{\vec{r}})\\
iF(r){\cal Y}_{j\tilde{l}m}(\hat{\vec{r}})
\end{array}
\right),
\end{equation}
where $\tilde{l}=2j-l=j\pm 1/2$ for $l=j\mp 1/2$, and 
\begin{equation}
{\cal Y}_{jlm}(\hat{\vec{r}})=\sum_{m_l,m_s}
\left.\left\langle l\,m_l\,\frac{1}{2}\,m_s\right| jm\right\rangle 
Y_{lm_l}(\hat{\vec{r}})\chi_{m_s},
\end{equation}
is the spin-angular wave function, 
$Y_{lm_l}$ and $\chi_{m_s}$ being the spherical harmonics and the 
spin wave function, respectively. 
The Dirac equation with a spherical vector potential $V(r)$ and a scalar 
potential $S(r)$ then reads\cite{S67,G97},
\begin{equation}
\left(
\begin{array}{cc}
U & \hbar c\left(-\frac{d}{dr}+\frac{\kappa}{r}\right) \\
\hbar c\left(\frac{d}{dr}+\frac{\kappa}{r}\right) & W-2mc^2
\end{array}
\right)
\left(
\begin{array}{c}
G \\ F
\end{array}
\right) 
=E
\left(
\begin{array}{c}
G \\ F
\end{array}
\right), 
\end{equation}
where $\kappa=\mp(j+1/2)$ for $j=l\pm1/2$, $U(r)=V(r)+S(r)$, and 
$W(r)=V(r)-S(r)$. 
It is easy to show that $G$ and $F$ are proportional to $r^{l+1}$ and 
$r^{\tilde{l}+1}$, respectively, around the origin, $r\sim 0$. 
Notice that $|\Psi^{n+1}\rangle$ satisfies 
$(H-E_0)|\Psi^{n+1}\rangle = (H-E_0+\Delta T)|\Psi^n\rangle$ (see 
Eq. (\ref{iteration2})). 
This implies that $G$ and $F$ at $T+\Delta T$ satisfy 
\begin{eqnarray}
&&\left(
\begin{array}{cc}
U-E_0 & \hbar c\left(-\frac{d}{dr}+\frac{\kappa}{r}\right) \\
\hbar c\left(\frac{d}{dr}+\frac{\kappa}{r}\right) & W-2mc^2-E_0
\end{array}
\right)
\left(
\begin{array}{c}
G^{n+1}(r) \\ F^{n+1}(r)
\end{array}
\right) \nonumber \\
&&\hspace{5cm}=\left(
\begin{array}{c}
\widetilde{G}^n(r) \\ \widetilde{F}^n(r)
\end{array}
\right), 
\label{radial}
\end{eqnarray}
where $\widetilde{G}$ and $\widetilde{F}$ are defined as 
\begin{eqnarray}
&&\left(
\begin{array}{c}
\widetilde{G}^n(r) \\ \widetilde{F}^n(r)
\end{array}
\right) \nonumber \\
&&\equiv
\left(
\begin{array}{cc}
U-E_0+\Delta T & \hbar c\left(-\frac{d}{dr}+\frac{\kappa}{r}\right) \\
\hbar c\left(\frac{d}{dr}+\frac{\kappa}{r}\right) & W-2mc^2-E_0+\Delta T
\end{array}
\right)
\left(
\begin{array}{c}
G^n(r) \\ F^n(r)
\end{array}
\right). \nonumber \\
\end{eqnarray}
We solve these equations by discretizing the radial coordinate with 
a spacing of $\Delta r$ and imposing the box boundary condition, that is, 
the wave functions vanish at the edge of the box. 
Denoting the wave functions at $r_k\equiv k\Delta r$ as $G_k$ and $F_k$, 
and using the three-point formula for the first derivative, 
$\psi'_k\sim(\psi_{k+1}-\psi_{k-1})/2\Delta r$, 
Eq. (\ref{radial}) reads 
\begin{equation}
\vec{\phi}_{k+1}+\vec{A}_k\vec{\phi}_k-\vec{\phi}_{k-1}=\vec{b}_k,
\label{recursion}
\end{equation}
with 
\begin{equation}
\vec{\phi}_k\equiv
\left(
\begin{array}{c}
G_k^{n+1} \\ F_k^{n+1}
\end{array}
\right),
~~~~~~
\vec{b}_k\equiv
\frac{2\Delta r}{\hbar c}\left(
\begin{array}{c}
\widetilde{F}_k^{n} \\ -\widetilde{G}_k^{n}
\end{array}
\right).
\end{equation}
Here, $\vec{A}_k$ is a 2$\times$2 matrix defined by 
\begin{equation}
\vec{A}_k\equiv \left(
\begin{array}{cc}
\frac{2\Delta r}{r_k}\kappa & 
\frac{2\Delta r}{\hbar c}(W_k-E_0-2mc^2) \\
-\frac{2\Delta r}{\hbar c}(U_k-E_0) & 
-\frac{2\Delta r}{r_k}\kappa 
\end{array}
\right). 
\end{equation}
We assume 
\begin{equation}
\vec{\phi}_{k+1}=\vec{\alpha}_k\vec{\phi}_k+\vec{\beta}_k, 
\label{wf}
\end{equation}
where $\vec{\alpha}_k$ is a 
2$\times$2 matrix and $\vec{\beta}_k$ is a two-component vector. 
Substituting this to Eq. (\ref{recursion}), one finds 
\begin{eqnarray}
\vec{\alpha}_{k-1}&=&(\vec{A}_k+\vec{\alpha}_k)^{-1}, \\
\vec{\beta}_{k-1}&=&(\vec{A}_k+\vec{\alpha}_k)^{-1}
(\vec{b}_k-\vec{\beta}_k). 
\end{eqnarray}
These equations can be solved inwards from 
$r_{N-1}$ with  
$\vec{\alpha}_{N-1}=\vec{\beta}_{N-1}=0$, which ensures 
that the wave functions vanish at $R_{\rm max}=r_N$. 
Once $\vec{\alpha}_k$ and $\vec{\beta}_k$ are so obtained, the 
wave functions $\vec{\phi}_k$ can be constructed with Eq. (\ref{wf}) 
outwards from $\vec{\phi}_{k=0}=0$. 

\begin{figure}[htb]
\includegraphics[clip,scale=0.5]{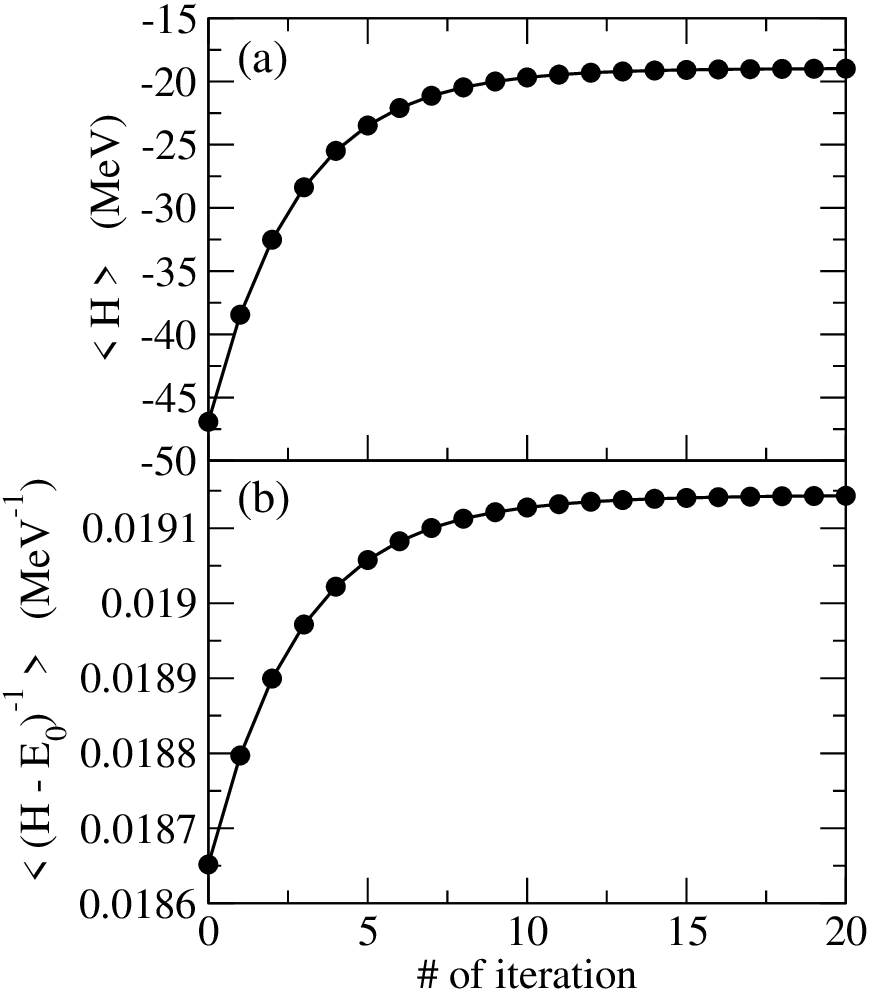}
\caption{
The expectation value of the Hamiltonian $H$ (Fig. 2(a)) and 
that of the inverse of 
the Hamiltonian, $1/(H-E_0)$, (Fig. 2(b)) at each iteration for the 
neutron 1p$_{1/2}$ state in $^{16}$O. 
$E_0$ is set to be the depth of the Woods-Saxon potential $U(r)=V(r)+S(r)$, 
and the step $\Delta T$ is taken to be 10 MeV. 
}
\end{figure}

Let us now numerically investigate the performance of the 
inverse Hamiltonian method. 
To this end, we use spherical Woods-Saxon potentials for $U(r)$ and $W(r)$ 
which correspond 
to $^{16}$O. The parameters of the Woods-Saxon potentials are 
taken from Ref. \cite{KR91}. 
Figure 2 shows the convergence feature for the neutron 1p$_{1/2}$ state 
($\kappa=1$), where 
the quantum numbers refer to the upper component of the wave function. 
The coordinate space is discretized 
up to $R_{\rm max}$=15 fm with $\Delta r=0.1$ fm. 
The energy shift $E_0$ is taken to be the depth of the potential $U(r)$, 
that is, $E_0=-71.28$ MeV, and the size of 
step $\Delta T$ is taken to be 10 MeV. 
For the initial wave functions, we take $F(r)=0$ and $G(r)$ to be the non-relativistic 
1p$_{1/2}$ wave function of a Woods-Saxon potential 
with $V_0=-51$ MeV, $R_0=1.27\times 16^{1/3}$ fm, and 
$a$=0.67 fm \cite{ZMR03}. 
Fig. 2(a) indicates that the energy of the 1p$_{1/2}$ state 
is quickly converged as the number of iteration increases. 
The converged value is $E=-18.974$ MeV, that is almost identical 
to the exact value, $E=-18.976$ MeV. 
Fig. 2(b) also shows that the expectation value of the 
inverse of the Hamiltonian,
\begin{equation}
\langle (H-E_0)^{-1}\rangle
=\frac{1}{\Delta T}\cdot\left\langle \Psi^n\left|
\left(1+\frac{\Delta T}{H-E_0}\right)\, 
\right|\Psi^n\right\rangle - \frac{1}{\Delta T},
\end{equation}
monotonically increases to the converged value, although it may not 
be the case for $\langle H \rangle$. 

\begin{figure}[htb]
\includegraphics[clip,scale=0.5]{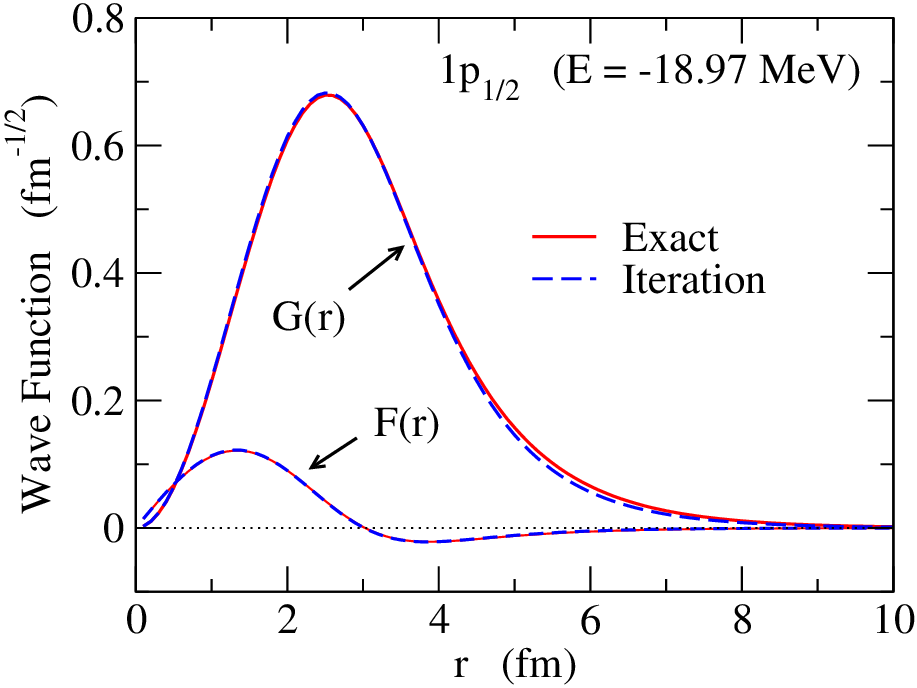}
\caption{(Color online)
Comparison of the wave function for the neutron 1p$_{1/2}$ state 
in $^{16}$O obtained with the inverse Hamiltonian method (the dashed lines) 
to the exact solutions (the solid lines). }
\end{figure}

Figure 3 shows the wave functions for the 1p$_{1/2}$ state. 
The solid lines show the exact solution of the Dirac equation obtained with 
the Runge-Kutta method, while the dashed lines show the wave function 
obtained with the inverse Hamiltonian method. 
One can clearly see that the iterative method well reproduces the 
exact wave function. 

We have checked that the performance of the inverse Hamiltonian method is 
good also for other states with different 
quantum numbers as well, although the accuracy seems 
to be somewhat improved 
if the energy shift $E_0$ can be chosen as close as possible to the 
exact value. Evidently, the inverse Hamiltonian method 
provides an efficient method to solve a Dirac equation iteratively. 

Before we close this paper, we would like to point out that 
the states in the Dirac sea can be also obtained 
within the same scheme by changing $\Delta T$ by $-\Delta T$ in 
Eq. (\ref{iteration2}). 
This is so because the highest energy state in the Dirac sea corresponds to 
the minimum of $1/(H-E_0)$ (see Fig. 1(b)). 
We have obtained the bound states in the Dirac sea in this way for $^{16}$O, 
and have confirmed that the exact wave functions are well reproduced. 

In summary, we have proposed a new scheme to iteratively solve 
a Dirac equation. The idea of the new scheme is to maximize the inverse 
of a shifted Hamiltonian, $1/(H-E_0)$, and thus we call it the inverse 
Hamiltonian method. By choosing $E_0$ to be between the highest energy 
in the Dirac sea and the lowest energy in the Fermi sea, the expectation 
value of $1/(H-E_0)$ monotonically converges to the exact value, 
$1/(E-E_0)$. 
The upper and the lower components of a wave function can be simultaneously 
obtained with this method in the coordinate space representation. 
We have applied this method to neutron states in $^{16}$O 
using spherical Woods-Saxon potentials. 
We have shown that the inverse Hamiltonian method efficiently reproduces the 
exact energies and wave functions for the states both in the Fermi 
and Dirac seas. 
In this paper, we have assumed spherical symmetry for nuclear potentials. 
It will be an interesting future work to apply this method to self-consistent 
relativistic mean-field calculations in three-dimensional space. 

\bigskip

This work was supported 
by the Grant-in-Aid for Scientific Research (C), Contract No. 
22540262 from the Japan Society for the Promotion of Science.

\end{document}